\documentclass[
aps,%
10pt,%
final,%
notitlepage,%
oneside,%
twocolumn,%
nobibnotes,%
nofootinbib,%
superscriptaddress,%
noshowpacs,%
centertags]%
{revtex4}

\begin{document}
\selectlanguage{english}

\title{Interstellar Plasma Turbulence Spectrum Toward the
Pulsars PSR~B0809$+$74 and B0950$+$08}

\author{\firstname{T.~V.} \surname{Smirnova}}

\author{\firstname{V.~I.} \surname{Shishov}}
\affiliation{Pushchino Radio Astronomy Observatory, Astro Space
Center of the Lebedev Institute of Physics, Russian Academy of
Sciences, Pushchino, Moscow oblast', 142290 Russia}

\begin{abstract}
Interstellar scintillations of pulsars PSR~B0809$+$74 and
B0950$+$08 have been studied using observations at low
frequencies (41, 62, 89, and 112~MHz). Characteristic temporal
and frequency scales of diffractive scintillations at these
frequencies have been determined. The comprehensive analysis of
the frequency and temporal structure functions reduced to the
same frequency has shown that the spectrum of interstellar plasma
inhomogeneities toward both pulsars is described by a power law.
The exponent of the spectrum of fluctuations of interstellar
plasma inhomogeneities toward PSR~B0950$+$08 ($n = 3.00\pm 0.05$)
appreciably differs from the Kolmogorov exponent. Toward
PSR~B0809$+$74 the spectrum is a power law with an exponent $n =
3.7\pm 0.1$. A strong angular refraction has been detected toward
PSR~B0950$+$08. The distribution of inhomogeneities along the
line of sight has been analyzed; it has been shown that the
scintillations of PSR~B0950$+$08 take place on a turbulent layer
with enhanced electron density, which is localized at
approximately 10~pc from the observer. For PSR~B0809$+$74 the
distribution of inhomogeneities is quasi-uniform. Mean-square
fluctuations of electron density on inhomogeneities with a
characteristic scale $\rho_0 = 10^7$~m toward four pulsars have
been estimated. On this scale the local turbulence level in the
10-pc layer is 20~times higher than in an extended region
responsible for the scintillations of PSR~B0809$+$74.
\end{abstract}

\maketitle

\section{INTRODUCTION}

Pulsars represent a good tool for the study of interstellar plasma
(ISP), because they possess very small angular sizes and intense
emission. The investigation of their intensity fluctuations in the
frequency--time domain allows us to study the ISP spectrum in
various directions of our Galaxy. As was shown in [1, 2], the
observational data on pulsar scintillations are statistically
well described by a power-law spectrum of inhomogeneities with an
exponent $n = 3.67$ (i.e., by the Kolmogorov spectrum) in a very
broad range of spatial scales $\rho$. However, a scatter in the
values of the spectrum exponent $n$ is observed for the same
value of $\rho$ for different sources; this testifies that in
local directions of the Galaxy the spectrum can differ from the
Kolmogorov spectrum. The application of a new method for the
study of ISP using a comprehensive analysis of a structure
function obtained from multifrequency observations of pulsar
scintillations has revealed a difference of the spectrum from
Kolmogorov's for PSR~B0329$+$54, B0437$-$47, and B1642$-$03
[3--5]. The purpose of this paper was a study of the ISP spectrum
ISP toward nearby and powerful at meter wavelengths pulsars
PSR~B0809$+$74 and B0950$+$08. The distances to these pulsars and
their velocities are known from parallax measurements [6]: $R =
433$~pc, $V = 102$~km/s and $R = 262$~pc, $V = 36.6$~km/s for
PSR~B0809$+$74 and B0950$+$08 respectively. Since they are the
nearest pulsars, the analysis of their scintillations allows us
to study the region of interstellar plasma nearest to the Sun.

The obtained earlier observational data demonstrate the presence
of three components of interstellar turbulent plasma in the solar
neighborhood at distances of the order of 1~kpc and less. The
first component is turbulent plasma with a statistically
quasi-uniform distribution in the space between spiral arms
(component A in the classification of [7, 8]). The second
component represents a cavern with a depleted electron density
($n_e\cong 0.005$~cm$^{-3}$) in the solar neighborhood with a
scale of about 200--300~pc perpendicularly to the Galactic plane
and 50--100~pc in the Galactic plane. This cavern has been
detected in X-ray observations [9]. The presence of a cavern with
a lowered turbulence level on a scale of the order of 100~pc has
been revealed also by the analysis of scintillations of pulsars
with dispersion measures from 3 to 35~pc~cm$^{-3}$ [10]. The
third component corresponds to a layer with an enhanced
turbulence level at a distance of about 10~pc from the Sun. This
layer has been detected in observations of interstellar
scintillations of quasars in the centimeter waveband [11, 12]. In
[5] this layer was classified as component C. In [5] it was also
shown that the pulsar PSR~J0437$-$4715 scintillates on
inhomogeneities of layer~C.

\begin{table}[h!]
\setcaptionwidth{\linewidth} \setcaptionmargin{0mm}
\onelinecaptionsfalse
\captionstyle{normal}  %
\caption{Parameters of the observations}
\begin{tabular}{l|c|c|c|c}
\hline \multicolumn{1}{c|} {Parameter}
 &  \multicolumn{4}{c}{$f$,~MHz} \\
\cline{2-5}
& 41 & \multicolumn{1}{c|}{62.43} & \multicolumn{1}{c|}{88.57} & \multicolumn{1}{c}{111.87} \\
\hline
\multicolumn{5}{c}{PSR~B0809$+$74} \\
$\Delta t$,~ms & 5.12; 2.56 & 5.15 & 5.38 & 2.56 \\
$N_{\textrm{ch}} $ & 128; 64\phantom{0} & 96 & 96 & 128 \\
$\Delta f$,~kHz & 1.25; 20\phantom{0.} & 20 & 20 & 20 \\
$T,$ s & 51.68 & 51.68 & 38.76 & 19.38 \\
\multicolumn{5}{c}{PSR~B0950$~+~$08} \\
$\Delta t$,~ms & 5.12; 2.56 & 2.56 & 2.56 & 2.56 \\
$N_{\textrm{ch}}  $ & 64\phantom{00.} & 96 & 128 & 128 \\
$\Delta f$,~kHz & 1.25; 20\phantom{0.} & 20 & 20 & 20 \\
$T$, s & 20.24 & 15.18 & 20.24 & 5.06 \\
\hline
\end{tabular}
\end{table}

\section{OBSERVATIONS AND DATA PREPROCESSING}

The observations of PSR~B0809$+$74 and B0950$+$08 were carried
out on the BSA and DKR-1000 radio telescopes of the Pushchino
Radio Astronomy Observatory (Astro Space Center, Lebedev
Institute of Physics, Russian Academy of Sciences) at frequencies
41, 62.43, 88.57~MHz (DKR-1000), and 111.87~MHz (BSA) in December
2001 -- January 2004. As the interference situation at low
frequencies was complex, we have used for the analysis only those
records where interference was small. We received linearly
polarized emission. Two multichannel receivers were used: a
128-channel receiver with a channel bandwidth $\Delta f = 20$~kHz
at frequencies 88.57, 62.43, and 111.87~MHz as well as a
128-channel receiver with a channel bandwidth $\Delta f =
1.25$~kHz at a frequency of 41~MHz. The time of observations on
BSA (frequency 111.87~MHz) in each session was 12 and 3.3~min for
PSR~B0809$+$74 and B0950$+$08 respectively. At low frequencies it
was 35.5~min for PSR~B0809$+$74 and 15.63~min for PSR~B0950$+$08.
Table~1 lists the time and frequency resolution ($\Delta t$ and
$\Delta f$) and the used number of channels ($N_{\textrm{ch}}$)
at each frequency. Individual pulses of the pulsars were recorded
on the computer disk in all channels with a period synchronized
with the precomputed topocentric pulse arrival time. Then the
signal in all channels was shifted in accordance with the
dispersion shift at the given frequency (reduction to the highest
frequency channel); in channels affected by interference (if any)
the signal was replaced by an average value found from the
adjacent channels. The amplification in all channels was reduced
to the same value by normalization, so that the noise dispersion
in all channels were equal to its average value found from
channels not affected by interference. At frequencies 41--89~MHz
the record was performed in a window 1.8\,$P_1$ ($P_1$ is the
pulsar period) and at 112~MHz in a window 0.9\,$P_1$. To improve
the signal-to-noise ratio ($S/N$), we averaged individual pulses:
from 15~pulses at 112~MHz to 40~pulses at low frequencies. The
averaging time $T$ for both pulsars at all frequencies is listed
in Table~1.

\section{CORRELATION ANALYSIS OF THE DATA}

To analyze the intensity variations of the pulse radiation as a
function of frequency (channel number) within the bandwidth of the
multichannel receiver, we formed spectra of individual pulses
$I(f_k)$ by averaging the signal in a selected range of
longitudes of the pulse and in a region outside the pulse, on the
noise segment of the record (for the same number of longitudes)
$I_N(f_k)$ for each channel. Here $f_k = f + (k-1) \Delta f$ is
the frequency of the $k$th channel, and $f$ is the frequency of
observation. Then we subtracted the noise component $I_N (f_k)$
from $I(f_k)$, and the spectra of individual pulses thus obtained
were analyzed with the purpose to get information in the
frequency as well as in the time domains.

\begin{figure}[h!]
\includegraphics[width=\columnwidth]{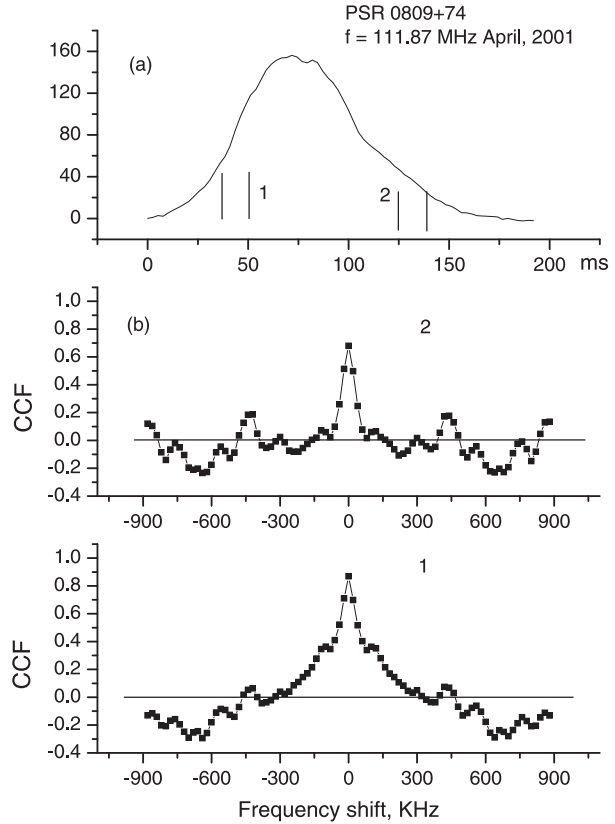}
\setcaptionmargin{5mm} \onelinecaptionsfalse \captionstyle{normal}
\caption{(a) Mean pulse profile of PSR~B0809$+$74 at a frequency
of 111.87~MHz as observed on April 6, 2001. Vertical axis:
intensity in relative (computer) units; horizontal axis: time
inside the pulse in milliseconds. Longitude ranges (interval
{\textit{1}} and interval {\textit{2}}) in which the pulse
intensity was averaged are marked. (b) Mean normalized CCF between
the spectra of adjacent pulses whose intensities were averaged in
the above-mentioned longitude ranges: {\textit{1}} (lower panel)
and {\textit{2}} (upper panel).}
\end{figure}

Characteristic frequency ($f_{\textrm{dif}}$) and temporal
($t_{\textrm{dif}}$) scales of scintillations were determined
using the correlation analysis. We computed the mean
cross-correlation function (CCF) between the spectra of adjacent,
noiseless pulses averaged in the given range of longitudes of the
pulse emission; CCF was normalized to the product of rms
deviations $\sigma_1\sigma_2$. We considered a pulse as noiseless
if its amplitude exceeded a level of 4$\sigma_N$. The mean
profile of the pulsar was derived by addition of all individual
pulses in the given observational session. Since the pulse
averaging time $T$ (Table~1) was considerably shorter than the
scintillation timescale, the decorrelation of the spectra of
adjacent (averaged) pulses was insignificant; in return, we
completely eliminated the uncorrelated component of noise. The
frequency scale $f_{\textrm{dif}}$ was determined as a frequency
shift at which CCF decreased by a factor of two.

Since we receive linearly polarized emission, we must take into
consideration the influence of polarization on its
frequency--time structure. At low frequencies the degree of
polarization for pulsars studied by us is high: about 60\% for
PSR~B0809$+$74 [13] at the longitudes of the leading part of the
mean profile and much lower in its trailing part; (70--80)\% for
PSR~B0950$+$08 [14] throughout the profile. The rotation measure
is $\textrm{RM} = {-}11.7$~rad/m$^2$ for PSR~B0809$+$74. The
frequency of the Faraday rotation is
\begin{gather}
P_F\textrm{~ [MHz]} = 17.475 f^3/{\textrm{RM}},
\label{eq:1}
\end{gather}
where $f$ is the observational frequency in hundreds of
megahertz. Accordingly, $P_F = 2140$~kHz at 112~MHz and $P_F =
59$~kHz at 41~MHz. The influence of polarization on the
correlation analysis for PSR~B0809$+$74 is shown in Fig.~1.
Longitude ranges in which we averaged the intensity to obtain the
spectra used in the calculation of the average CCF are shown.
When averaging the leading part of the pulse (interval
{\textit{1}}), a superposition of two structures---narrow-band
(produced by propagation of the radiation in the interstellar
plasma) and broadband (due to a much slower intensity variation
with frequency as a result of the rotation of the polarization
plane across the receiver bandwidth)---is visible. When averaging
the trailing part (interval {\textit{2}}), the slow component is
virtually absent; this confirms the small degree of polarization
at these longitudes. In the frequency--time data analysis of
PSR~B0809$+$74 we have used the average over the longitudes of
the trailing part of the mean profile.

\begin{figure}[h!]
\includegraphics[width=\columnwidth]{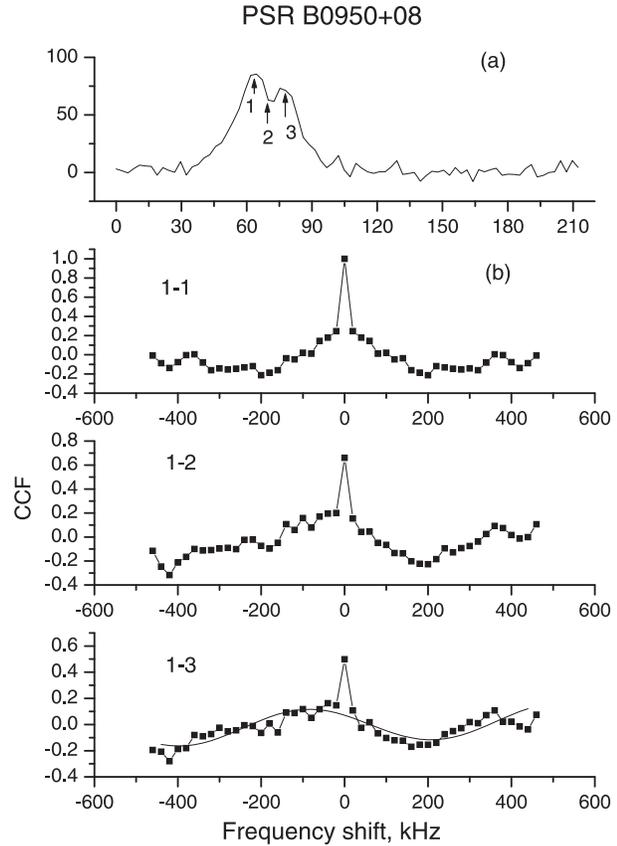}
\setcaptionmargin{5mm} \onelinecaptionsfalse \captionstyle{normal}
\caption{(a) Mean profile of PSR~B0950$+$08 at a frequency of
41~MHz as observed on January 16, 2004. Vertical axis: intensity
in relative (computer) units; horizontal axis: time inside the
pulse in milliseconds. The time resolution is 2.56~ms, frequency
resolution is 20~kHz. Arrows show longitudes at which the pulse
intensity was selected in all channels for the calculation of the
correlation functions. (b) Mean CCF between the pulse spectra
taken at longitudes {\textit{1}}--{\textit{1}},
{\textit{1}}--{\textit{2}}, and {\textit{1}}--{\textit{3}}. The
curve in the lowermost graph is the fitted sine wave with a
frequency of 600~kHz.}
\end{figure}

\begin{figure}[h!]
\includegraphics[height=\columnwidth,angle=270]{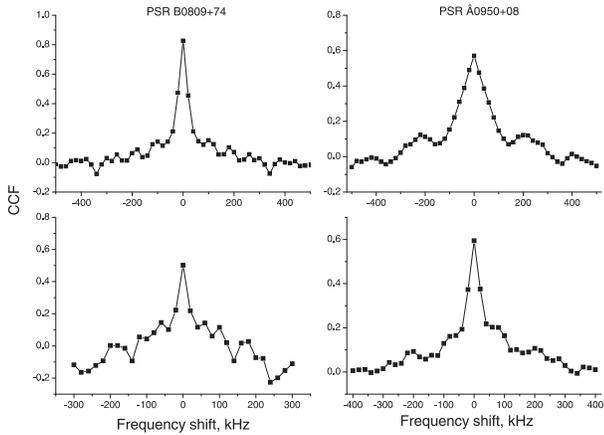}
\setcaptionmargin{5mm} \onelinecaptionsfalse \captionstyle{normal}
\caption{Mean normalized CCF for the spectra of adjacent pulses at
 88.57~MHz (upper graphs) and 62.43~MHz (lower graphs) for
PSR~B0809$+$74 (left) and PSR~B0950$+$08 (right).}
\end{figure}

\begin{figure}[h!]
\includegraphics[height=\columnwidth,angle=270]{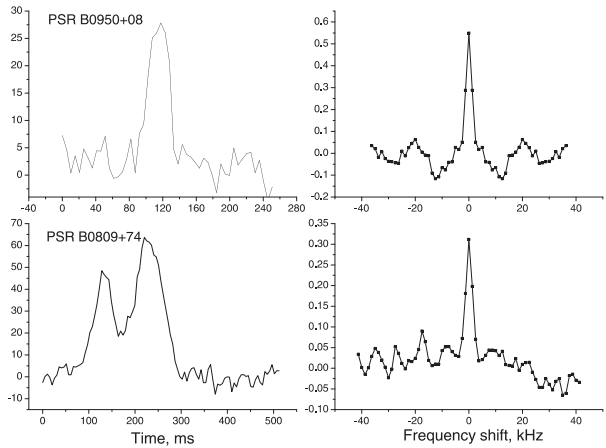}
\setcaptionmargin{5mm} \onelinecaptionsfalse \captionstyle{normal}
\caption{Mean profiles of PSR~B0950$+$08 and PSR~B0809$+$74 at
41~MHz as observed on December 25, 2003 (left) and mean normalized
CCF between the spectra of adjacent pulses on this frequency
(right). The time resolution is 5.12~ms, and the frequency
resolution 1.25~kHz. In the left graphs, the vertical axis plots
the intensity in relative (computer) units, and the horizontal
axis the time inside the pulse in milliseconds.}
\end{figure}

\begin{figure}[h!]
\includegraphics[width=5cm,angle=270]{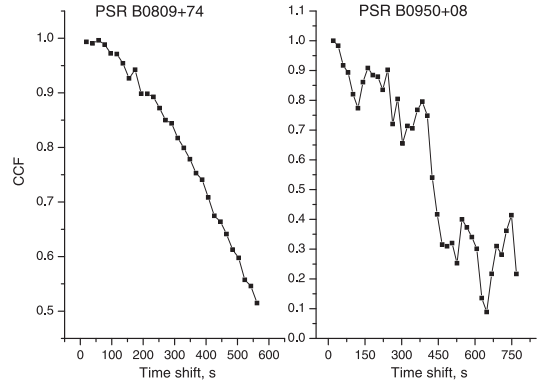}
\setcaptionmargin{5mm} \onelinecaptionsfalse \captionstyle{normal}
\caption{Mean normalized CCF for the spectra of pulses taken at
appropriate time intervals (horizontal axis); left: for
PSR~B0809$+$74 at 111.87~MHz, right: for PSR~B0950$+$08 at 
88.57~MHz.}
\end{figure}

For PSR~B0950$+$08 the rotation measure is considerably smaller
than for PSR~B0809$+$74; however, in the literature its values
appreciably differ: in the catalog of pulsars [15] $\textrm{RM} =
(1.35 \pm 0.15)$~rad/m$^2$, in [14] $\textrm{RM} =
(4\textrm{--}6)$~rad/m$^2$, and in [16] $\textrm{RM} = ({-} 0.66
\pm 0.04)$~rad/m$^2$. Probably, authors [14] underestimated the
contribution of the ionosphere to the obtained rotation measure.
For $\textrm{RM} = 1$~rad/m$^2$ the Faraday rotation frequency is
$P_F = 20.4$~MHz at 112~MHz and $P_F = 1200$~kHz at 41~MHz. The
influence of polarization on our observations at 41~MHz is shown
in Fig.~2. The profile averaged over the session (Fig.~2a) is
double-peaked with a separation between the components of 13~ms.
Arrows shown longitudes at which we selected pulse intensities
over all channels to form the spectrum at the given longitude.
Then we calculated mean normalized CCF between the spectra at
longitudes {\textit{1}}--{\textit{1}} (autocorrelation function,
ACF), {\textit{1}}--{\textit{2}}, and {\textit{1}}--{\textit{3}}
for all noiseless pulses. Figure~2b shows mean CCF between the
spectra at the corresponding longitudes. In these observations we
used the receiver with a channel bandwidth of 20~kHz. The narrow
unresolved feature corresponding to the frequency scale of
diffractive scintillations at 41~MHz has a maximum at the zero
frequency shift, while the slow component is shifted to the left
with an increase in the longitude offset of the spectra; this is
due to the change in the position angle across the pulsar profile.
Fitting a sine wave to the slow CCF component
{\textit{1}}--{\textit{3}} yields the modulation frequency $P_F =
(600\pm 60)$~kHz, which corresponds to $\textrm{RM} = (2\pm 0.2)
$~rad/m$^2$. The observations were carried out in the nighttime,
and we suppose that the contribution of the ionosphere to this
value was not more than 1~rad/m$^2$. Consequently, RM toward
PSR~B0950$+$08 is about 1~rad/m$^2$, and the effect of
polarization on the frequency--time structure of the emission for
this pulsar is considerably smaller than for PSR~B0809$+$74.
Therefore, for obtaining the spectra we used the average over all
longitudes determined by the level of 0.25 of the mean profile
maximum of the pulsar.

\begin{table}[h!]
\setcaptionwidth{\linewidth} \setcaptionmargin{0mm}
\onelinecaptionsfalse
\captionstyle{normal}  %
\caption{Characteristic scales of diffractive scintillations}
\begin{tabular}{l|c|c|c|c}
\hline
  \multicolumn{1}{c|}{Parameter} &  \multicolumn{4}{c}{$f$,~MHz} \\
\cline{2-5}
& 41 & 62.43 & 88.57 & 111.87 \\
\hline
\multicolumn{5}{c}{PSR~B0809$+$74} \\
$f_{\textrm{dif}},$~kHz & \phantom{0.}$2 \pm 0.6$ & $7 \pm 3$ & $20 \pm 10$ & $45 \pm 5$\phantom{0} \\
$t_{\textrm{dif}},$~s & -- & -- & $450 \pm 70$\phantom{0} & $600 \pm 100$ \\
\multicolumn{5}{c}{PSR~B0950$~+~$08} \\
$f_{\textrm{dif}},$~kHz & $1.5 \pm 0.4$  & $25 \pm 10$ & $100 \pm 40$\phantom{0} & $220 \pm 60$\phantom{0} \\
$t_{\textrm{dif}},$~s & $350 \pm 100$  & $400 \pm 100$ & $450 \pm 120$ & ${>}200$ \\
\hline
\end{tabular}
\end{table}

Figure~3 shows for both pulsars mean CCF between the spectra of
adjacent pulses at 88.57~MHz (upper graphs) and 62.43~MHz (lower
graphs) obtained in individual observational sessions. Table~2
lists, together with their errors, the average values of
characteristic scales of scintillations $f_{\textrm{dif}}$ found
from all sessions. The errors correspond to standard deviations
from the mean. The slow CCF component is due to the effect of
polarization. It is visible that with decreasing frequency the
relative amplitude of the slow component increases; this
testifies to an increase in the degree of polarization with
frequency. Figure~4 presents the mean profiles of PSR~B0809$+$74
and B0950$+$08 for one of the observational sessions at the low
frequency 41~MHz (left-hand graphs) and mean CCF between the
spectra of adjacent pulses at this frequency (right-hand graphs).
The frequency resolution here was 1.25~kHz for both pulsars. To
improve the signal-to-noise ratio for PSR~B0950$+$08 observed at
$f = 41$~MHz with the narrow-band receiver, we used a time
constant of 10~ms; therefore, the components of the mean profile
merge. For PSR~B0809$+$74 a well-resolved two-peak profile with a
separation between components of 96~ms is observed.

We determined the characteristic timescale of scintillations as
the time shift at which the coefficient of correlation between
the spectra separated by an appropriate time interval decreases
by a factor of two. Figure~5 presents the corresponding functions
for both pulsars obtained from observations at one of the
frequencies: 111.87~MHz for PSR~B0809$+$74 (left) and 88.57~MHz
for PSR~B0950$+$08 (right). The average values of the
characteristic scintillation scales $t_{\textrm{dif}}$ found from
all sessions are given together with their errors in Table~2. The
errors correspond to standard deviations from the mean. For
PSR~B0809$+$74 we did not manage to determine $t_{\textrm{dif}}$
at 41 and 62~MHz because of a poor signal-to-noise ratio.

\section{STRUCTURAL ANALYSIS OF THE DATA}

We used the obtained frequency--time correlation functions for
the structural analysis of the data and, accordingly, for the
study of the spectrum of ISP inhomogeneities. The power-law
spectrum of electron density fluctuations is defined as
\begin{gather}
\Phi_{Ne} (q) = C_{Ne}^2 q^{-n},
\label{eq2}
\end{gather}
where $C_{Ne}^2$ characterizes the plasma turbulence level, $q =
2\pi/\rho$ and $\rho$ are, respectively, the spatial frequency
and spatial scale of an inhomogeneity in the plane perpendicular
to the line of sight (generally it is a three-dimensional
vector). The structure function of phase fluctuations $D(\rho)$
and spectrum $\Phi_{Ne}$ are related by the Fourier transform. In
the case of a statistically uniform distribution of
inhomogeneities in the medium $D(\rho)$ is described by
relationship [17]
\begin{gather}
\label{eq3}
D_s (\rho) = (k \Theta_0 \rho)^{n-2}, \\
\nonumber (k\Theta_0)^{n-2} = A (n) (\lambda r_e)^2 C_{Ne}^2 L/(n-1), \\
\nonumber A (n) = \frac{2^{4-n} \pi^3}{[\Gamma^2 (n/2) \sin (\pi n/2)]}.
\end{gather}
Here $\Theta_0$ is the characteristic scattering angle, $\lambda$
is the wavelength, $k = 2\pi/\lambda$ is the wavenumber, $r_e$ is
the electron classical radius, $L$ is the effective distance to
the turbulent layer or distance to the pulsar in the case of a
uniform distribution of turbulence along the line of sight. The
spatial scale is related to the timescale by a simple
relationship $\rho = V\Delta t$, where $V$ is the velocity of
motion of the line of sight across the picture plane. If the
pulsar velocity considerably exceeds the velocity of the observer
and of the motion of the turbulent medium, then it determines the
displacement of the line of sight. As was shown in [3], for small
time shifts $\Delta t$ the phase structure function $D_s(\Delta
t)$ can be obtained from the correlation function of intensity
fluctuations $B_I(\Delta t)$:
\begin{gather}
D_s (\Delta t) = {{B_I (0) -B_I (\Delta t)} \over {\langle
I\rangle^2}} \quad {\textrm{at}} \quad \Delta t \leq t_{\textrm{dif}},
\label{eq4}
\end{gather}
where $\langle I\rangle$ is the mean intensity for an
observational session. In the frequency domain we have used the
relationship
\begin{gather}
D_s (\Delta f) = {{B_I (0) -B_I (\Delta f)} \over{\langle
I\rangle^2}} \quad {\textrm{for}} \quad\Delta f \leq
f_{\textrm{dif}},
\label{eq5}
\end{gather}
where $\Delta f$ is the frequency shift. To reduce the data from
different frequencies to a single frequency $f_0$, it is
necessary to scale temporal and frequency structure functions in
accordance with the law:
 \begin{gather}
D_s (f_0, \Delta t (f_0), \Delta f (f_0)) = D_s (f, \Delta t,
\Delta f) (f/f_0)^2. \label{eq6}
\end{gather}
As noted in [3], $\Delta t (f_0) = \Delta t (f)$; however,
$\Delta f(f)\neq\Delta f(f_0)$. In the case of purely diffractive
scintillations we have
\begin{gather}
\Delta f_d (f_0) = (f_0/f)^2 \Delta f (f).
\label{eq7}
\end{gather}
In the presence of a strong angular refraction, i.e., when the
refraction angle considerably exceeds the scattering angle, the
frequency dependence $\Delta f$ is quite different:
\begin{gather}
\Delta f_r (f_0) = (f_0/f)^3 \Delta f (f).
\label{eq8}
\end{gather}
We have used these formulas in the analysis of our data
considering two models, diffractive and refractive.

\begin{figure}[h!]
\includegraphics[width=\columnwidth]{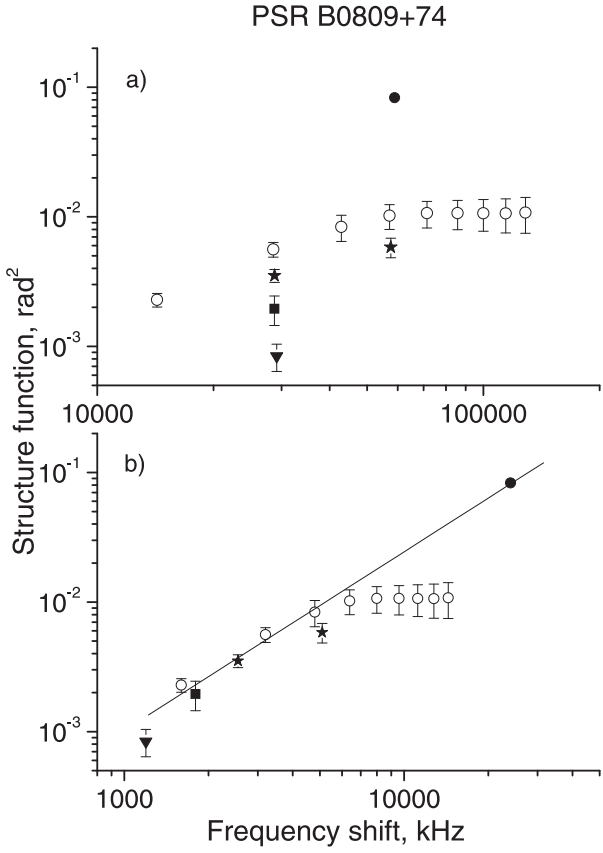}
\setcaptionmargin{5mm} \onelinecaptionsfalse \captionstyle{normal}
\caption{Frequency structure functions of phase fluctuations for
PSR~B0809$+$74 in two models: refractive (a) and diffractive (b).
Open circles: 111.87-MHz data; asterisks: 88.57~MHz; square:
62.43~MHz; triangle: 41~MHz; filled circle: 408~MHz [18]. The data
for all frequencies have been reduced to the same frequency $f_0 =
1000$~MHz. The straight line least-square-fitted to the first
points of structure functions for the diffractive model has a
slope $\beta = 1.3 \pm 0.2$.}
\end{figure}

\begin{figure}[h!]
\includegraphics[height=\columnwidth,angle=270]{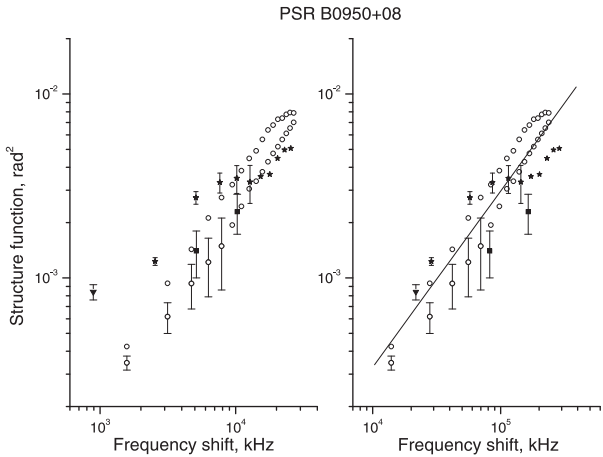}
\setcaptionmargin{5mm} \onelinecaptionstrue \captionstyle{normal}
\caption{Same as in Fig.~6, but for PSR~B0950$+$08. The fitted
straight line has a slope $\beta = 0.96 \pm 0.05$.}
\end{figure}

\begin{figure}[h!]
\includegraphics[height=\columnwidth,angle=270]{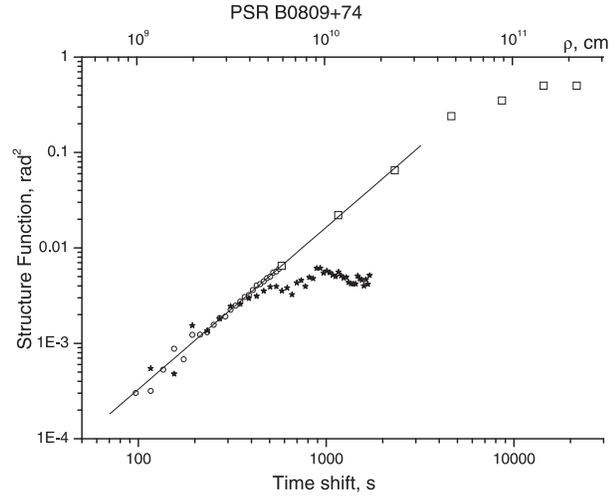}
\setcaptionmargin{5mm} \onelinecaptionsfalse \captionstyle{normal}
\caption{Temporal structure function of phase fluctuations for
PSR~B0809$+$74. Open circles: 111.87~MHz; asterisks: 88.57~MHz;
small open squares: 933~MHz [19]. The data for all frequencies
have been reduced to the same frequency $f_0 = 1000$~MHz. The
straight line has been least-square-fitted to the initial points
of the structure functions. Top: axis of the corresponding spatial
scales.}
\end{figure}

\begin{figure}[h!]
\includegraphics[height=\columnwidth,angle=270]{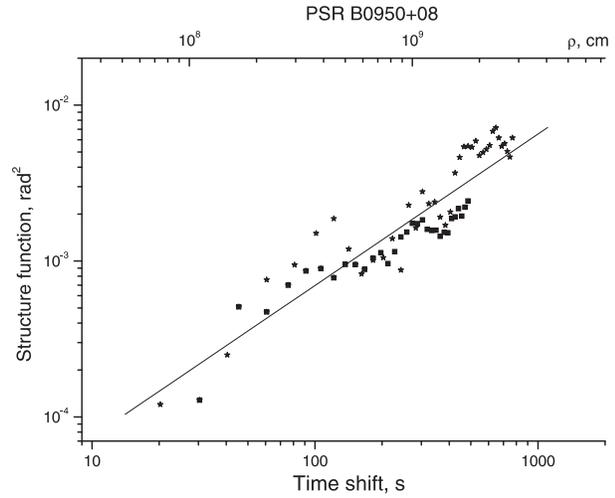}
\setcaptionmargin{5mm} \onelinecaptionsfalse \captionstyle{normal}
\caption{Temporal structure function of phase fluctuations for
PSR~B0950$+$08. Asterisks: 88.57 MHz; small filled squares:
62.43~MHz. The data for all frequencies have been reduced to the
same frequency $f_0 = 1000$~MHz. The straight line has been
least-square-fitted. Top: axis of the corresponding spatial
scales.}
\end{figure}

Figures~6 and 7 show for both models frequency structure
functions of phase fluctuations in the double logarithmic scale
for PSR~B0809$+$74 and PSR~B0950$+$08 respectively. All the data
were reduced to the same frequency $f_0 = 1000$~MHz. As in our
previous publication on the study of ISP spectra toward a number
of pulsars [3--5], we have used this value of $f_0$ for
convenience of a comprehensive analysis of all the data. In the
figures the data at different frequencies are shown by different
symbols. In Fig.~6 a filled circle marks a point obtained from
observations at 408~MHz [18]. Statistical errors of the structure
functions were estimated using equation (B12) from [19]. From
Fig.~6 it is visible that for PSR~B0809$+$74 the diffractive model
(Fig.~6b) describes the experimental data much better than the
refractive model (Fig.~6a). A straight line least-square-fitted
to the first points of the structure functions has a slope $\beta
= 1.3 \pm 0.2$. For fitting we must take the values at shifts
smaller than or of the order of the characteristic frequency
scale of scintillations. For PSR~B0950$+$08 (Fig.~7) the data at
different frequencies in both models differ not so strongly as
for PSR~B0809$+$74; however, the scatter of the points for the
refractive model (Fig.~7a) is smaller. The largest difference
takes place for a point obtained from observations at 41~MHz
(filled triangle). The slope of the fitted straight line is
$\beta = 0.96\pm 0.05$.

Figures~8 and 9 show temporal structure functions of phase
fluctuations for PSR~B0809$+$74 and PSR~B0950$+$08 respectively.
In addition to our data, we have used in Fig.~8 also a structure
function obtained in [19] from observations at 933~MHz (open
squares). It is visible that the data at different frequencies
for PSR~B0809$+$74 presented in the double logarithmic scale are
well described by a unified power-law spectrum. A straight line
fitted to the experimental points has a slope $\alpha = 1.7\pm
0.04$. As it is visible from Fig.~9, for PSR~B0950$+$08 temporal
structure functions at different frequencies reduced to 1000~MHz
satisfy a power law with an exponent $\alpha = 1.0 \pm 0.05$. The
scale of the upper horizontal axis in Fig.~8 and 9 corresponds to
spatial scales of interstellar plasma inhomogeneities $\rho = V
\Delta t$. Here we have used the pulsar velocities measured in
[6]: $V = 102$ and 36.6~km/s for PSR~B0809$+$74 and
PSR~B0950$+$08 respectively.

\begin{figure}[h!]
\includegraphics[height=\columnwidth,angle=270]{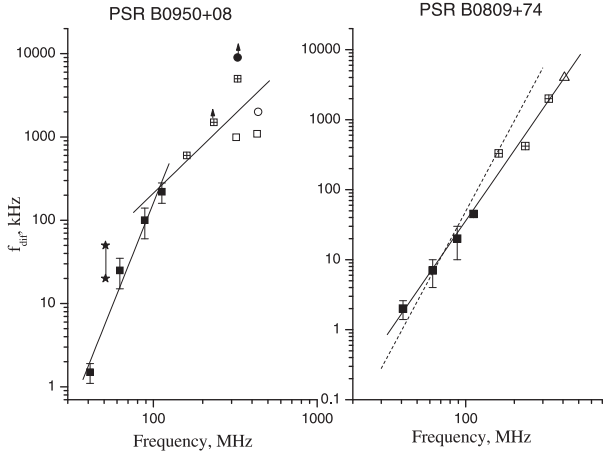}
\setcaptionmargin{5mm} \onelinecaptionsfalse \captionstyle{normal}
\caption{Characteristic frequency scale of scintillations as a
function of the frequency of observation for PSR~B0950$+$08 (left)
and PSR~B0809$+$74 (right). The filled squares show the scales
obtained in this work, the asterisks the data of [20], the crust
the data of [21], the open circle the data of [22], the filled
circle the data of [10], the triangle the data of [18], and the
open squares: data from [7]. The lines are the result of
least-squares fits. The slope for PSR~B0950$+$08 using our data is
$\gamma = 4.9 \pm 0.6$, while the slope for the data at
frequencies ${\geq} 112$~MHz is $\gamma = 2 \pm 0.6$. The fitting
for PSR~B0809$+$74 fitting using all the points yielded the slope
of $\gamma = 3.4 \pm 0.15$. The dashed line with a slope of 4.3 is
the expected $f_{\textrm{dif}}(f)$ dependence.}
\end{figure}

\begin{figure}[h!]
\includegraphics[width=\columnwidth]{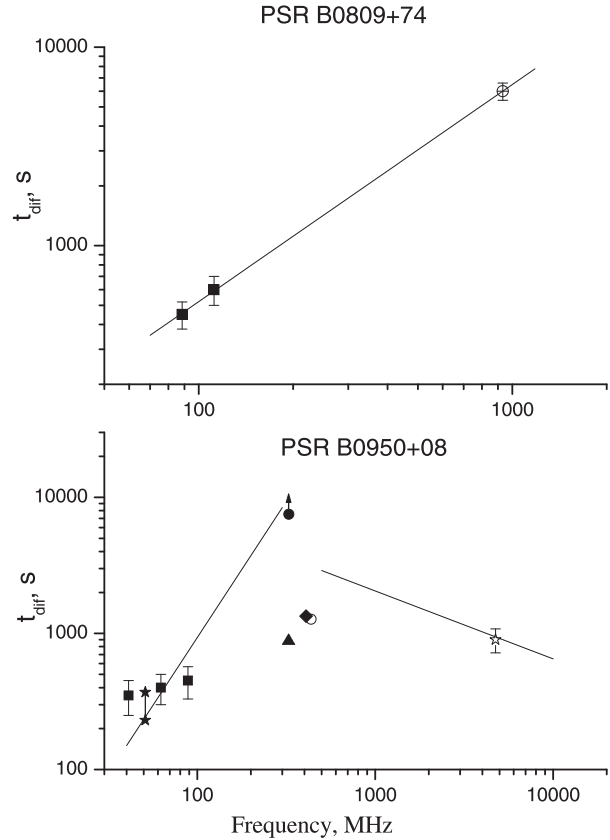}
\setcaptionmargin{5mm} \onelinecaptionsfalse \captionstyle{normal}
\caption{Characteristic temporal scale of scintillations as a
function of the frequency of observation; top: PSR~B0809$+$74,
bottom: PSR~B0950$+$08. Filled squares: scales obtained in this
work, triangle: the data from [23], filled asterisks: [20], the
open asterisk: [24], the diamond the data of [25], open circle:
[19]. A straight line has been least-square-fitted to the data for
PSR~B0809$+$74; its slope is 1.1. Asymptotic straight lines with
slopes of $\alpha = 2$ and $\alpha = {-} 0.5$ corresponding to
equations (10) and (14) are shown for PSR~B0950$+$08.}
\end{figure}

\section{DISCUSSION}

As shown above, the temporal structure functions for
PSR~B0809$+$74 and PSR~B0950$+$08 are described by a power law,
and, accordingly, the spectrum of interstellar plasma
inhomogeneities toward these pulsars is a power-law one with an
exponent $n = \alpha + 2$; this corresponds to $n = 3.0 \pm 0.05$
for PSR~B0950$+$08 and $n = 3.7 \pm 0.04$ for PSR~B0809$+$74.
Whereas for PSR~B0809$+$74 the measured turbulence spectrum
matches to the Kolmogorov form ($n = 3.67$), for PSR~B0950$+$08
it is much flatter. In addition, we can say that toward
PSR~B0950$+$08 an appreciable angular refraction takes place. As
shown in [3], for the diffractive model the slope of the
frequency structure function is related to that of the temporal
structure function as $\beta = \alpha /2$, and for the refractive
model $\beta = \alpha$. For PSR~B0950$+$08 we have
$\beta\approx\alpha$. For PSR~B0809$+$74 the accuracy of the
determination of $\beta$ is low, and within the $3\sigma$ error
limits the relationship $\beta = \alpha /2$ is fulfilled, though
a more convincing evidence of the absence of angular refraction
in this direction is a strong deviation of the data on the
frequency structural function from this model (Fig.~6).

Figures~10 and 11 present the characteristic frequency and
temporal scales of scintillations as functions of the frequency of
observation for both pulsars. Filled squares show our
measurements, other symbols denote the data taken from [7, 10,
18--25]. For the reduction of the scales $f_d$ obtained by a
different method in [21] to our measurements, we have used the
relationship $f_{\textrm{dif}} = 0.3f_d$. We have
least-square-fitted straight lines to the observation data. For
PSR~B0950$+$08 (Fig.~10) fitting was done separately to the
low-frequency data of this work (slope $\gamma = 4.9 \pm 0.6$),
and to the data at frequencies $f> 112$~MHz (slope $\gamma = 2 \pm
0.6$). For PSR~B0809$+$74 fitting for all points (Fig.~10) yields
$\gamma = 3.4 \pm 0.15$. As shown above, the frequency CCF of the
flux fluctuations of the pulsar PSR~B0809$+$74 is described by
the model of diffractive scintillations on inhomogeneities of the
power-law spectrum of interstellar plasma turbulence with an
exponent $n = 3.7$. For the diffractive model the characteristic
frequency scale of scintillations is [17]
\begin{gather}
f_{\textrm{dif}} = c/(\pi R\Theta_0)^2 \propto f^{2n/(n-2)},
\label{eq9}
\end{gather}
where $c$ is the velocity of light. For $n = 3.7$ the expected for
PSR~B0809$+$74 value $\gamma = 4.3$ is shown in Fig.~10 with a
dotted line. This line is also consistent with the experimental
points, except for the high-frequency ones. It is possible that
the values of $f_{\textrm{dif}}$ measured at high frequencies are
underestimated because of the limited total frequency band of the
observation. The characteristic timescale of the scintillations
for the diffractive model is
\begin{gather}
t_{\textrm{dif}} \approx 1/ k \Theta_0 V \propto f^{2/(n-2)},
\label{eq10}
\end{gather}
where $k$ is the wavenumber and $V$ is the velocity of motion of
the line of sight with respect to the turbulent medium. For $n =
3.7$ the expected for PSR~B0809$+$74 value $2/(n-2) = 1.18$; this
agrees well with the experimentally obtained figure 1.1 (Fig.11).
In [26] it was shown that the scintillation parameters for the
pulsar PSR~B0809$+$74 and quasar B0917$+$624, which is located
close to the pulsar on the celestial sphere, are determined by the
same turbulent medium and this medium is distributed almost
uniformly on a scale of the order of 500~pc.

For the pulsar PSR~B0950$+$08 the temporal and frequency
correlation functions of the flux fluctuations are described by a
more sophisticated model.

Firstly, we should take into account that the observational data
correspond to two modes of scintillations. At high frequencies
($f> f_{\textrm{cr}}$) scintillations are weak, and at low
frequencies ($f < f_{\textrm{cr}}$) the observation data
correspond to the diffractive component of strong scintillations.
Near the critical frequency $f_{\textrm{cr}}$ scintillations
correspond to the focusing mode, which has been poorly studied,
either theoretically or experimentally. According to our data,
the scintillations index is about unity at all observational
frequencies; thus, scintillations are strong at frequencies
$f\leq 112$~MHz. Consequently, the critical frequency
$f_{\textrm{cr}} > 112$~MHz. According to the data [25] obtained
at $f = 4.8$~GHz, scintillations of the pulsar PSR~B0950$+$08 are
weak. In the same paper an estimate of the critical frequency
$f_{\textrm{cr}} < 700$~MHz is given. In [27] on the basis of the
compilation of the observational data of 1970s an estimate of the
critical frequency $f_{\textrm{cr}}\approx 300$~MHz is given.

Secondly, it is necessary to take into account the presence of a
strong angular refraction. In this case, the characteristic
frequency scale of diffractive (strong) scintillations is
described by the relationship [3]
\begin{gather}
\label{eq11} f_{\textrm{dif}} \approx c/ \pi
r_{\textrm{ef}}\Theta_0 \Theta_{\textrm{ref}} \propto f^{2 +
n/(n-2)}, \\
\nonumber f < f_{\textrm{cr}},
\end{gather}
where $r_{\textrm{ef}}$ is the effective distance to the turbulent
layer, $\Theta_{\textrm{ref}}$ is the characteristic angle of
refraction, $\Theta_0$ is the characteristic scattering angle
determined by (3). For $n = 3$ equation (11) yields
$f_{\textrm{dif}}\propto f^5$. The observational data indeed
correspond to a power law with a turnover. At low frequencies ($f
\leq 112$~MHz) the exponent $\gamma = 4.9 \pm 0.6$; this is
consistent with the expected value for the mode of strong
scintillations in the presence of refraction. Therefore, we can
assert that the critical frequency $f_{\textrm{cr}} > 112$~MHz. In
the case of weak scintillations we should replace the scattering
angle $\Theta_0$ with the Fresnel angle
\begin{gather}
\Theta_{\textrm{Fr}} = (1/ kr_{\textrm{ef}})^{1/2}.
\label{eq12}
 \end{gather}
As a result we obtain
\begin{gather}
f_{\textrm{dif}} \approx c/ \pi
r_{\textrm{ef}}\Theta_{\textrm{Fr}}\Theta_{\textrm{ref}}\propto
f^{2.5}. \label{eq13}
\end{gather}
At high frequencies $f\geq 112$~MHz the exponent $\gamma = 2 \pm
0.6$; this also agrees with the expected value for the mode of
weak scintillations.

The characteristic timescale of scintillations for PSR~B0950$+$08
is also described by a power law with a turnover. At low
frequencies scintillations are strong, and the characteristic
timescale of scintillations is determined by formula (10). At high
frequencies scintillations are weak, and the characteristic
timescale of scintillations is
\begin{gather}
t_{\textrm{dif}} \approx (r_{\textrm{ef}}/k)^{1/2}
/V_{\textrm{ef}}\propto f^{-1/2}, \label{eq14}
\end{gather}
where $V_{\textrm{ef}}$ is the velocity of motion of the line of
sight with respect to the turbulent layer. The asymptotic
relationships (10) and (14) are shown in Fig.~11 with straight
lines. We see that the theoretical dependences agree well enough
with the experimental points. From the timescale of weak
scintillations we can determine the distance to the layer
responsible for the radiowave modulation. If $L$ is the distance
from the observer to the turbulent layer and $R$ is the distance
from the observer to the pulsar, the effective distance is
\begin{gather}
1/r_{\textrm{ef}} = 1/(R - L) + 1/L.
\label{eq15}
\end{gather}
The effective velocity is
\begin{gather}
V_{\textrm{ef}} = V_{\textrm{o}} (R - L) /R + V_pL/R,
 \label{eq16}
\end{gather}
where $V_{\textrm{o}}$ and $V_{\textrm{p}}$ are the components of
the velocities of the observer and pulsar in the picture plane
respectively. The pulsar velocity is $V_p = 36.6 $~km/s [6];
therefore, $V_p\cong V_{\textrm{o}}$, and $V_{\textrm{ef}}\cong
V_p \cong 30$~km/s. Using this value of $V_{\textrm{ef}}$ in (14)
and supposing $t_0 = 15$~min at the frequency $f = 4.8$~GHz [25],
we obtain
\begin{gather}
L (R - L) /R \cong 9.7\textrm{~pc}.
\label{eq17}
\end{gather}
For this equation we find the solutions $L_1 = 10$~pc and $L_2 =
252$~pc. Therefore, the turbulent layer is either near the
observer or near the pulsar. It should be noted that parameters
of the scintillations of the pulsar PSR~B0950$+$08 are rather
close to those of the pulsar B0437$-$47 [5], which is located at
approximately the same galactic longitude. For the latter pulsar
it was shown in [5] that its scintillations as well as those of
the quasar PKS 0405$-$385, at a small angular distance, are
determined by a layer of the medium with enhanced turbulence at
10~pc from the observer. Therefore, we can accept that the pulsar
PSR~B0950$+$08 scintillates on the same layer of the turbulent
medium and that the right solution is $L_1 = 10$~pc.

\begin{table}[h!]
\setcaptionwidth{\linewidth} \setcaptionmargin{0mm}
\onelinecaptionstrue \captionstyle{flushleft} \caption{Local
turbulence levels}
\begin{tabular}{ccccc}
\hline PSR & $D_s(\rho_0)$, &
 $L$, &
 $D_s(\rho_0)/L$, &
 $\Delta N_e$,  \\
 & rad$^2$ &  pc & rad$^2$/pc & cm$^{-3}$ \\ \hline
 B0329$+$54 & 0.034 & 1000 & $3\times 10^{-5}$ & $7.7\times 10^{-4}$ \\
 B0437$-$47 & 0.0033 & 10 & $3\times 10^{-4}$ & $2.4\times 10^{-3}$ \\
 B0809$+$74 & 0.0003 & 433 & $1\times 10^{-6}$ & $1.4\times 10^{-4}$ \\
 B0950$+$08 & 0.002 & 10 & $2\times 10^{-4}$ & $2\times 10^{-3}$ \\
\hline
\end{tabular}
\end{table}

Let us estimate the turbulence level toward PSR~B0809$+$74 and
PSR~B0950$+$08. Note that we cannot use for the comparative
analysis the values of $C_{Ne}^2$, because they change their
dimension with changing spectrum exponent $n$. We will use the
values of the structure function on a selected scale $\rho_0 =
10^7$~m at a frequency $f = 1$~GHz. The local level of turbulence
is related to the gradient $D_s (\rho_0)$ as follows: $(d/L) D_s
(\rho_0) \approx D_s (\rho_0) /L$, where $L$ is the characteristic
thickness of the layer of the medium. The value of $D_s
(\rho_0)/L$ is proportional to the mean square of electron density
fluctuations on inhomogeneities with a characteristic scale
$\rho_0$; for $\rho_0 = 10^7$~m we have:
 \begin{gather}
\label{eq18}
D_s (\rho_0) /L \textrm{~[rad$^2$/pc]} \\
\nonumber {} \cong 50 \langle [\Delta N_e (\rho_0)]^2 \rangle
\textrm{~ [~cm$^{-6}$]}.
\end{gather}
Table~3 lists the values of $D_s (\rho_0)$ at a frequency $f =
1$~GHz together with the values of $D_s (\rho_0)/L$ and $\Delta
N_e$ for the pulsars studied by us in [3, 5] and in this paper.
The distribution of the estimated local turbulence levels
corresponds to three components of the interstellar medium
mentioned in the Introduction. The greatest values of $\Delta N_e$
correspond to the layer with an enhanced turbulence level
(layer~C): the pulsars PSR~B0437$-$47 and PSR~B0950$+$08
scintillate on its inhomogeneities. The minimum value of $\Delta
N_e$ obtained toward the pulsar PSR~B0809$+$74 corresponds to a
cavern. The intermediate value of $\Delta N_e$ is obtained toward
the pulsar PSR~B0329$+$54; it corresponds to turbulent plasma in
the space between spiral arms (component~A).

\section{CONCLUSION}

As a result of the analysis of the phase structure functions
obtained by us from the study of interstellar scintillations
toward PSR~B0809$+$74 and PSR~B0950$+$08 at low frequencies
41--112~MHz with invoking the previously published higher
frequency data, we have shown that the spectrum of interstellar
plasma inhomogeneities toward both pulsars is described by a
power law. The spectrum exponent toward PSR~B0950$+$08 differs
appreciably from the Kolmogorov exponent; it is $n = 3.00 \pm
0.05$. Toward PSR~B0809$+$74 the spectrum is a power law with an
exponent $n = 3.7 \pm 0.1$; within the error limits this
correspond to the Kolmogorov spectrum. The analysis of the
frequency dependence of the diffraction parameters has shown that
toward PSR~B0950$+$08 the transition to the weak scintillation
mode takes place approximately at a frequency of 200--300~MHz;
here is a turnover in relationships $t_{\textrm{dif}}(f)$ and
$f_{\textrm{dif}}(f)$. The experimental data match well enough
the dependence expected by the theory.

As a result of the analysis of the temporal and frequency phase
structural functions for a purely diffractive model and model with
a strong angular refraction, we have shown that toward
PSR~B0950$+$08 there exists a strong angular refraction. The
conducted analysis of scintillations of this pulsar at different
frequencies has shown that the distribution of inhomogeneities
along the line of sight is not uniform and that the
scintillations of PSR~B0950$+$08 take place on a turbulent layer
with enhanced electron density (layer C), which is located at a
distance of about 10~pc from the observer. The same layer is
responsible for the scintillations of PSR~B0437$-$47 and and of
the quasar PKS~0405$-$385. Here the local turbulence level on a
scale $\rho_0 = 10^7$~m is 20 times higher than in the extended
region with a size of the order of 430~pc responsible for the
scintillations of PSR~B0809$+$74.

\section*{ACKNOWLEDGMENTS}

This work was supported by the Russian Foundation for Basic
Research (project codes 06-02-16810 and 06-02-16888) and by the
Program of Basic Research of the Presidium of the Russian Academy
of Sciences ``Origin and evolution of stars and galaxies''.

\end{document}